\documentclass[aps,prb,twocolumn,superscriptaddress,floatfix]{revtex4}
\usepackage{graphicx}
\usepackage{afterpage}
\usepackage{dcolumn}
\usepackage{bm}
\newcommand{\glue}{$\tilde{\Pi}(\omega)$ }

\newcommand{\hg}{HgBa$_{2}$CuO$_{4+\delta}$ }
\newcommand{\bitri}{Bi$_{2}$Sr$_{2}$Ca$_{2}$Cu$_{3}$O$_{10+\delta}$ }
\newcommand{\biduo}{Bi$_{2}$Sr$_{2}$CaCu$_{2}$O$_{8+\delta}$ }
\newcommand{\bimono}{Bi$_{2}$Sr$_{2}$Cu$_{2}$O$_{6+\delta}$ }

\begin{document}
\preprint{APS/123-QED}
\title{Optical determination of the relation between the electron-boson coupling function and the critical temperature in high T$_c$ cuprates.}
\author{E. van Heumen}
\author{E. Muhlethaler}
\author{A.B. Kuzmenko}
\affiliation{D\'epartement de Physique de la Mati\`ere
Condens\'ee, Universit\'e de Gen\`eve, quai Ernest-Ansermet 24,
CH1211 , Gen\`eve 4, Switzerland}
\author{H. Eisaki}
\affiliation{Nanoelectronics Research Institute, National
Institute of Advanced Industrial Science and Technology,
Tsukuba, Japan}
\author{W. Meevasana}
\author{M. Greven}
\affiliation{Department of Physics, Applied Physics, and
Stanford Synchrotron Radiation Laboratory, Stanford University,
Stanford, CA 94305}
\author{D. van der Marel}
\affiliation{D\'epartement de Physique de la Mati\`ere
Condens\'ee, Universit\'e de Gen\`eve, quai Ernest-Ansermet 24,
CH1211 , Gen\`eve 4, Switzerland}
\begin{abstract}
We take advantage of the connection between the free carrier
optical conductivity and the glue function in the normal state,
to reconstruct from the infrared optical conductivity the
glue-spectrum of ten different high-T$_c$ cuprates revealing a
robust peak in the 50-60 meV range and a broad continuum at
higher energies for all measured charge carrier concentrations
and temperatures up to 290 K. We observe that the strong
coupling formalism accounts fully for the known strong
temperature dependence of the optical spectra of the high
T$_{c}$ cuprates, except for strongly underdoped samples. We
observe a correlation between the doping trend of the
experimental glue spectra and the critical temperature. The
data obtained on the overdoped side of the phase diagram
conclusively excludes the electron-phonon coupling as the main
source of superconducting pairing.
\end{abstract}
\maketitle

\section{Introduction.}
The theoretical approaches to the high T$_c$ pairing mechanism
in the cuprates are divided in two main groups: According to
the first electrons form pairs due to a retarded attractive
interaction mediated by virtual bosonic excitations in the
solid\cite{scalapino-PRB-1986,varma-PRL-1989,millis-PRB-1990,dolgov-physc-1991,abanov-spec-2001}.
These bosons can be lattice vibrations, fluctuations of
spin-polarization, electric polarization or charge density. The
second group of theories concentrates on a pairing-mechanism
entirely due to the non-retarded Coulomb
interaction\cite{anderson-sci-2007} or so-called
Mottness\cite{phillips-annphys-2006}. Indeed, optical
experiments have found indications for mixing of high and low
energy degrees of freedom when the sample enters into the
superconducting
state\cite{basov-sci-1999,molegraaf-science-2002,carbone-PRB-2006a,carbone-PRB-2006b}.

An indication that both mechanisms are present was obtained by
Maier, Poilblanc and Scalapino\cite{maier-PRL-2008}, who showed
that the 'anomalous' self-energy associated with the pairing
has a small but finite contribution extending to an energy as
high as $U$, demonstrating that the pairing-interaction is, in
part, non-retarded. The experimental search for a pairing glue
will play an essential role in determining the origin of the
pairing interaction. Aforementioned glue is expressed as a
spectral density of these bosons, indicated as
$\alpha^2F(\omega)$ for phonons and $I^2\chi(\omega)$ for spin
fluctuations, here represented as the general, dimensionless
function $\tilde{\Pi}(\omega)$. An important consequence of the
electron-boson coupling is, that the energy of the
quasi-particles relative to the Fermi level, $\xi$, is
renormalized, and their lifetime becomes limited by inelastic
decay processes involving the emission of bosons. The
corresponding energy shift and the inverse lifetime, {\em i.e.}
the real and imaginary parts of the self-energy, are expressed
as the convolution of the 'glue-function' $\tilde{\Pi}(\omega)$
with a kernel $K(\xi,\omega,T)$ describing the thermal
excitations of the glue and the electrons\cite{kernel}
\begin{equation}
\Sigma(\xi)=\int K(\xi,\omega,T) \tilde{\Pi}(\omega)d\omega
  \label{sigma}
\end{equation}
In the absence of a glue and of scattering off impurities the
effect of applying an AC electric field to the electron gas is
to induce a purely reactive current response, characterized by
the imaginary optical conductivity
$4\pi\sigma(\omega)=i\omega_p^2/\omega$, where the plasma
frequency, $\omega_p$, is given by the (partial) f-sum rule for
the conduction electrons. The effect of coupling the electrons
to bosonic excitations is revealed by a finite, frequency
dependent dissipation, which can be understood as arising from
processes whereby a photon is absorbed by the simultaneous
creation of an electron-hole pair and a boson. As a result, the
expression for the optical conductivity in the normal state,
\begin{equation}
4\pi\sigma(\omega)=\frac{i\omega_p^2}{\omega+\hat{M}(\omega)},
\end{equation}
now contains a memory function or optical self
energy\cite{gotze-PRB-1972,timusk-footnote}. A particularly
useful aspect of this representation is that $\hat{M}(\omega)$
follows in a straightforward way from the experimental optical
conductivity. The optical self-energy is related to the single
particle self-energies by the expression\cite{pballen-PRB-1971}
\begin{equation}
\frac{\hat{M}(\omega)}{\omega}=
\left\{
 {\int \frac{f(\xi)-f(\xi+\omega)}{\omega+\Sigma^*(\xi)-\Sigma(\xi+\omega)}d\xi}\right\}^{-1}
 - 1
  \label{Kubo}
\end{equation}

The central assumption in the above is the validity of the
Landau Fermi-liquid picture for the normal state. The
aforementioned strong coupling analysis is therefore expected
to work best on the overdoped side of the cuprate phase
diagram, where the state of matter appears to become
increasingly Fermi liquid like. If antiferromagnetism is
necessary to obtain the insulating state in the undoped parent
compounds, as has been argued based on the doping trends of the
Drude spectral weight\cite{millis-nphys-2008}, the strong
coupling analysis may in principle be relevant for the entire
doping range studied. However, in the limit of strong
interactions aforementioned formalism needs to be extended,
{\em e.g.} with vertex corrections, and it eventually breaks down.
We therefore {\em define} the function \glue as the {\em
effective} quantity which, in combination with Eqs. \ref{sigma}
and \ref{Kubo}, returns the exact value of $\hat{M}(\omega)$
for each frequency. Defined in this way \glue captures {\em
all} correlation effects regardless whether the system is a
Fermi-liquid or not. This becomes increasingly relevant when
the doping is lowered below optimal doping.

Here we take advantage of the connection between the
temperature and frequency dependent conductivity in the normal
state and the glue-spectrum to test experimentally the
consequences of the standard approach, to check the internal
consistency of it, and to determine the range of doping where
internal consistency is obtained. For a d-wave superconductor,
the momentum dependence is essential to understand the details
of the pairing.  This, of course, is difficult to handle for
optical spectroscopy which is inherently a momentum integrated
probe. Nevertheless, optical spectra provide the important
information on the energy scale of the bosons involved and on
the doping and temperature evolution. The paper is organized as follows. In section \ref{holstein} we show that the temperature dependence of the optical spectra of the cuprates is well described within the strong coupling formalism described above. In section \ref{gluefunctions} we present the \glue functions for 10 different cuprates. These \glue functions are used in section \ref{Tc} to estimate critical temperatures and section \ref{implications} discusses the implications of these results with regard to the pairing mechanism in the cuprates. Finally, in section \ref{conclusion} we summarize our results.

\section{Internal consistency check of the strong coupling formalism.}\label{holstein}

In order to test whether the strong coupling analysis is
applicable to the cuprates we start with an important test of
its internal consistency: (i) we invert the data at 290 K to
obtain $\tilde{\Pi}(\omega)$, (ii) we use this $\tilde{\Pi}(\omega, 290 K)$ to \textit{predict} the optical spectra at lower temperatures. If the prediction faithfully reproduces the experimental spectra at these temperatures, we have a strong indication that the electronic structure and its evolution as a function of temperature are to a good approximation within the realm of strong coupling theory. We use a standard least squares routine to fit a histogram representation of \glue to our experimental
infrared spectra (see Appendix). The quantity \glue is shown in Fig. \ref{fig_selfenergy} for optimally doped \hg (Hg-1201)\cite{heumen-PRB-2007} for $T= 290$ K, together with
the optical self energies calculated from this function at
three different temperatures. For 290 K the theoretical curve
runs through the data points, reflecting the full convergence
of the numerical fitting routine.

It is interesting to notice, that the shoulder at 80 meV in the 100 K experimental data is reproduced by the same \glue function as the one used to fit the 290 K data.  It can be excluded that this shoulder is due to the pseudo-gap, since a gap is certainly absent for temperatures as high as 290 K. The shoulder is therefore entirely due to coupling of the electrons to a mode at approximately 60 meV. On the other hand, the considerable sharpening of this feature for temperatures lower than 100 K finds a natural explanation in the opening of a gap, as illustrated in the inset of Fig. \ref{fig_selfenergy}. We see that for 100 K the theoretical prediction also runs through the experimental data points. In other words, the strong temperature dependence of the experimental optical spectra is entirely due to the Fermi and Bose factors of Eqs. \ref{sigma} and \ref{Kubo}.

This example confirms the close correspondence between the features in $\hat{M}(\omega)$ and in $\tilde{\Pi}(\omega)$ pointed out in Ref. [\onlinecite{norman-PRB-2006}]. In particular the broad maximum in $\hat{M}(\omega)$ has its counterpart in the high intensity region of \glue terminating at 290 meV.
\begin{figure}[t]
\centering
\includegraphics[width=8.5 cm]{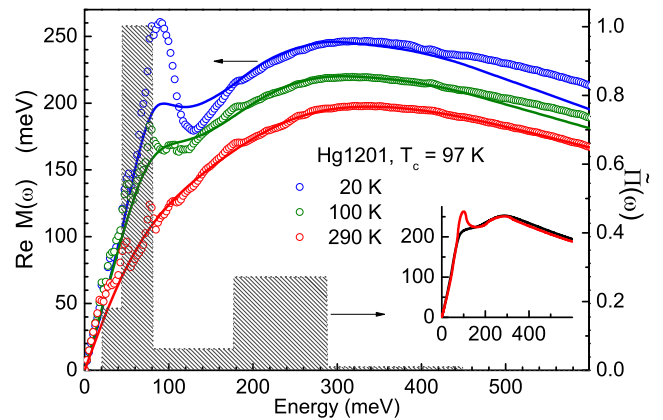}
\caption{Experimental optical self energy of \hg for 3 selected temperatures
(open circles). The solid curve at 290 Kelvin is obtained from a fit of \glue,
shown as the dashed surface. The solid curves at 100 K and 20 K were calculated
with the same \glue function corresponding to 290 Kelvin. This proves that the
self energy feature between 80 and 100 meV (a shoulder at 100 K and a peak at 20 K)
is caused by the prominent peak in \glue at approximately 60 meV. The sharpening
of this feature at low temperature is due to the superconducting gap,
an aspect not captured by Eq. \ref{Kubo} and therefore not reproduced in
the calculated solid curves. In the inset the gap-induced sharpening is
illustrated by the optical self energy without (black) and with (red) a 15 meV
superconducting gap, calculated using Allen's relation\cite{pballen-PRB-1971}.}
\label{fig_selfenergy}
\end{figure}
The internal consistency is therefore demonstrated by the fact that the large temperature dependence of the optical spectra is fully explained by the strong coupling formalism. This consistency was obtained for all samples, except for the most strongly underdoped single layer Bi2201 sample.

\section{Electron boson coupling function.}\label{gluefunctions}
\begin{figure*}[!t]
\centering
\includegraphics[width=17 cm]{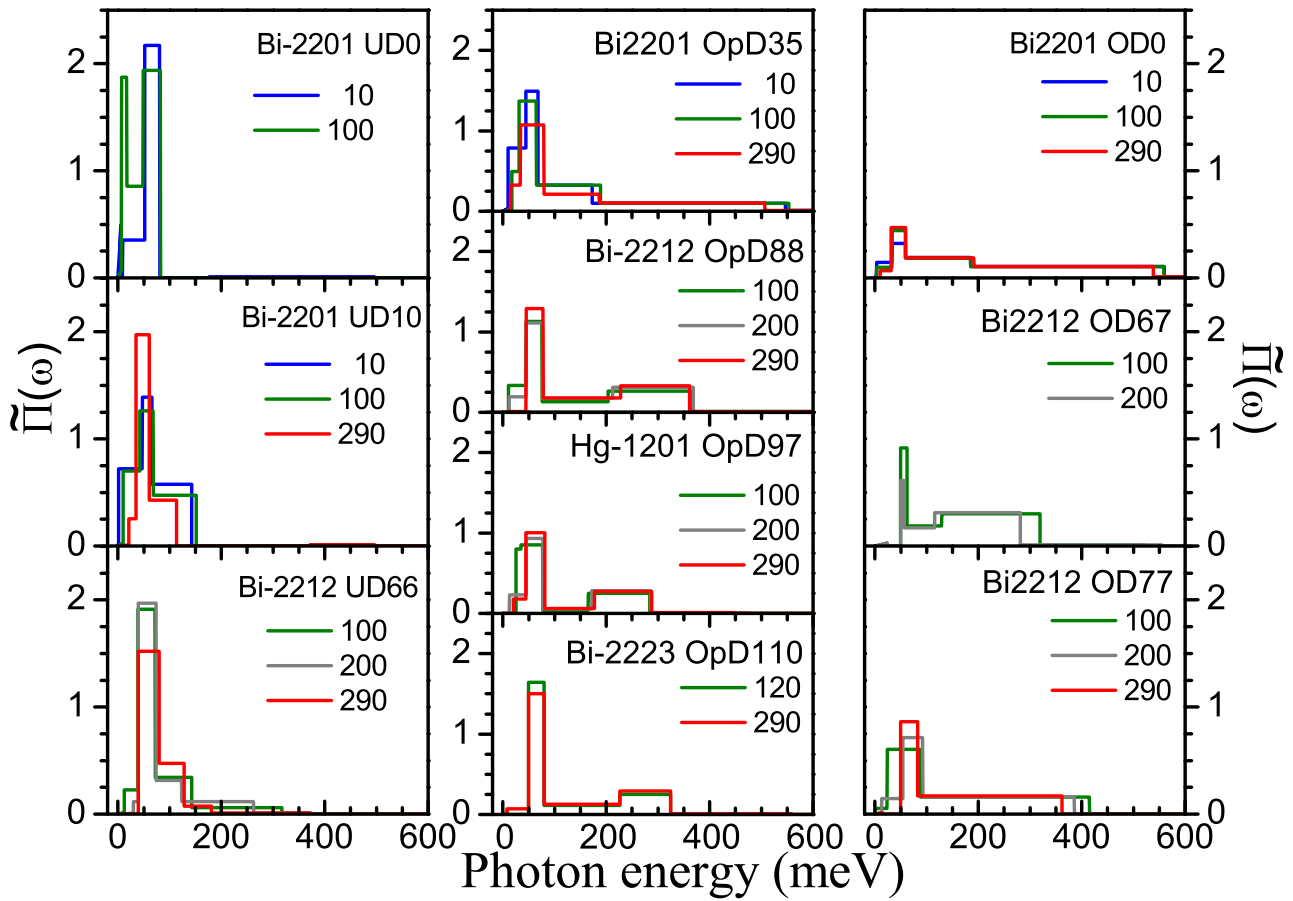}
\caption{Electron-boson coupling function \glue for Bi-2201 at
four different charge carrier concentrations (10 K, 100 K, 290
K), Bi-2212 at four charge carrier concentrations, and
optimally doped Bi-2223 and Hg-1201 (100 K, 200 K, 290 K). The
samples are ordered from underdoped to overdoped (left to
right) and low to high T$_{c}$ (top to bottom)}
\label{fig_glue}
\end{figure*}
As summarized in Fig. \ref{fig_glue}, we have analyzed
previously published optical spectra of 10 different samples
belonging to different families of materials, {\em i.e.}
optimally doped Hg-1201\cite{heumen-PRB-2007} and \bitri
(Bi-2223)\cite{carbone-PRB-2006b}, as well as four  \biduo
(Bi-2212) crystals
\cite{molegraaf-science-2002,carbone-PRB-2006a} with different
hole concentrations. In addition, we analyzed data for four
\bimono (Bi-2201) crystals with different hole
concentrations\cite{heumen-NJP-2009}.

Excellent fits were obtained for all temperatures, but the
\glue spectra exhibit a significant temperature dependence, in
particular at the low frequency side of the \glue spectrum.
Since all thermal factors contained in Eqs. \ref{sigma} and
\ref{Kubo} are, in principle, folded out by our procedure, the
remaining temperature dependence of \glue reflects the thermal
properties of the 'glue-function' itself. Such temperature
dependence is a direct consequence of the peculiar DC and far
infrared conductivity, in particular the $T$-linear DC
resistivity and $\omega/T$ scaling of $T\sigma(\omega,T)$ at
optimal doping\cite{dirk-nature-2003}. For the highest doping
levels both \glue and its temperature dependence diminish,
which is an indication that a Fermi liquid regime is
approached. The most strongly underdoped sample, Bi-2201-UD0,
exhibits an upturn of the imaginary part of the experimental
optical self-energy for $\omega\rightarrow 0$. This aspect of
the data can not be reproduced by the strong coupling
expression, resulting in an artificial and unphysical peak at
$\omega\approx 0$ of the fitted \glue function.

We observe two main features in the glue-function: A robust
peak at 50-60 meV and a broad continuum. The upper limit of
\glue is situated around approximately 300 meV for optimally
doped single layer Hg1201, and for the bilayer and trilayer
samples. The continuum extends to the highest energies (550 meV for the single-layer samples and 400 meV for
the bilayer) for the weakly overdoped samples, whereas the
continuum of the strongly doped bilayer sample extends to only
300 meV. There is also a clear trend of a contraction of the
continuum to lower energies when the carrier concentration is
reduced. Hence, part of the glue function has an energy well
above the upper limit of the phonon frequencies in the cuprates
($\sim$ 100 meV). Consequently the high energy part of \glue
reflects in one way or another the strong coupling between the
electrons themselves.

The most prominent feature, present in all spectra reproduced
in Fig. \ref{fig_glue}, is a peak corresponding to an average
frequency of 60 $\pm$ 3 meV at room temperature (see Appendix
for an estimate of the error bar). Perhaps the most striking
aspect of this peak is the fact that its energy is practically
independent of temperature (up to room temperature) and sample
composition. Moreover, the intensity and width are essentially
temperature independent. While our results confirm by and large
the observations of Hwang \textit{et al.} in the pseudo-gap
phase\cite{timusk4,timusk5}, the persistence of the 50-60 meV
peak to room temperature has not been reported before for these compounds. However, Collins {\em et al} obtained excellent fits to their infrared data of YBa$_2$Cu$_3$O$_7$ at 100 K and 250K using for both temperatures the same $\alpha^2F(\omega)$ spectrum with a peak at $\sim 35$ meV and a continuum extending up to 300 meV.  The 50-60 meV peak which we observe, arises most likely from the same boson that is responsible
for the 'kink' seen in angle resolved photoemission (ARPES)
experiments along the nodal direction in k-space at
approximately the same
energy\cite{bogdanov-PRL-2000,lanzara-NAT-2001,non-PRL-2006}.
The peak-dip-hump structure in the tunneling spectra
(STS)\cite{lee-nat-2006,levy-condmat-2007,zasad-PRL-2001} has
also been reported at approximately the same energy.

\section{Critical temperatures}\label{Tc}
\begin{table}[!t]
\begin{center}
\begin{tabular}{l|l|llll|llll|l|l}
\hline
x&&0.09&0.11&0.16&0.22&0.11&0.16&0.20&0.21&0.16&0.16\\
\hline
$T_c$&K&0&10&35&0&66&88&77&67&110&97\\
\hline
$\hbar\omega_p$ & eV         &1.75&1.77&1.92&1.93&2.36&2.35&2.45&2.33&2.43&2.10\\
$\hbar\tilde{\omega}$ &meV   & -  &70&81&103&92&124&116&154&101&81\\
$\lambda$         &      & -  &2.96&2.95&1.42&2.66&2.15&1.50&0.97&2.18&1.85\\
\hline
$\lambda_{pk}$         &      & -  &2.85&2.47&0.95&2.36&1.53&1.07&0.35&1.75&1.5\\
$\lambda_{cnt}$         &      & -  &0.11&0.48&0.47&0.3&0.62&0.44&0.62&0.43&0.35\\
$T_{c,pk}$&K&-&160&140&64&169&123&90&22&132&110\\
$T_{c,cnt}$&K&-&5&116&113&26&184&101&154&101&64\\
\hline \hline
\end{tabular}
\end{center}
\caption{\label{sampleparameters} Strong coupling parameters of
the ten compounds. The hole-doping is indicated on the first
line. From left to right: Bi2201 (columns 1 to 4), Bi2212
(columns 5 to 8), Bi2223 (9th column) and Hg1201 (column 10).
On rows 6-9 we indicate the partial coupling constants
$T_{c}$'s obtained when the \glue spectra are separated in a
contribution from the peak (pk, $\omega\le$ 100 meV) and from
the continuum (cnt, $\omega\ge$ 100) meV. All values are listed
for room temperature.} \label{table1}
\end{table}
One of the most important issues in the field of high T$_{c}$ is the question whether pairing is caused by the exchange of virtual bosons. These processes are described by a bosonic density of states function closely related to \glue. If the electron-electron interaction occurs uniquely in the $d$-wave channel, the superconducting critical temperature follows from the usual relation
\begin{equation}
T_c = 0.83 \tilde{\omega}
\exp(-(1+\lambda_{d})/\lambda_{d}),
\label{tc}
\end{equation}
where $\lambda_{d}$ is the coupling constant in the $d$-wave pairing channel is
\begin{equation}
\lambda_{d}=2\int_{0}^{\infty}\tilde{\Pi}_{d}(\omega)/\omega d\omega,
\end{equation}
and $\tilde{\Pi}_{d}(\omega)$ is the d-wave electron-boson coupling function. The effective frequency of the bosons responsible for the pairing interaction is obtained by taking the average of $\ln(\omega)$ weighted by electron-boson coupling function\cite{carbotte},
\begin{equation}
\ln(\tilde{\omega})=2\lambda^{-1}\int_0^{\infty}
\omega^{-1}\tilde{\Pi}_{d}(\omega)\ln(\omega)d\omega.
\end{equation}
To apply Eq. \ref{tc} to our experimentally measured \glue, we would need to map this function on the $d$-wave pairing channel. Boson fluctuations below a certain critical frequency act as pair breakers, as has been shown by Millis, Varma and Sachdev\cite{millis-PRB-1988} in the case of spin-fluctuation-mediated d-wave superconductivity. Clearly, it is not possible to separate pair-breaking from pair-forming contributions to \glue in an unambiguous way. To proceed we assume that the \textit{full} \glue function contributes favorably to the pairing. This means that our results overestimate the critical temperature. In Table \ref{table1} we indicate the total coupling constant, $\lambda$, and logarithmic frequency, $\tilde{\omega}$, for the room temperature \glue spectra. The coupling strength shows a strong and systematic increase with decreasing hole concentration, which probably requires a theoretical treatment beyond the strong coupling expansion. At the same time we see that $\tilde{\omega}$ shows the opposite trend.

An estimate of $T_c$, using the experimental values indicated
in Table \ref{table1}, gives values in the 100-200 K range.
The critical temperature can also be calculated straightforwardly from the s-wave Eliashberg equations\cite{owen-phys-1971} when \glue is known. As shown in Fig. \ref{fig_tc}, the $T_c$'s are in the 150-300 K range,
and they correlate with the experimentally observed doping
trends of $T_c$. The dome-shaped trend in the calculation is a
consequence of the increasing energy scale of
$\tilde{\Pi}(\omega)$ and the decreasing overall coupling
constant as a function of doping.

\section{Implications for the pairing mechanism.}\label{implications}
\begin{figure}[t!]
\centering
\includegraphics[width=8.5 cm]{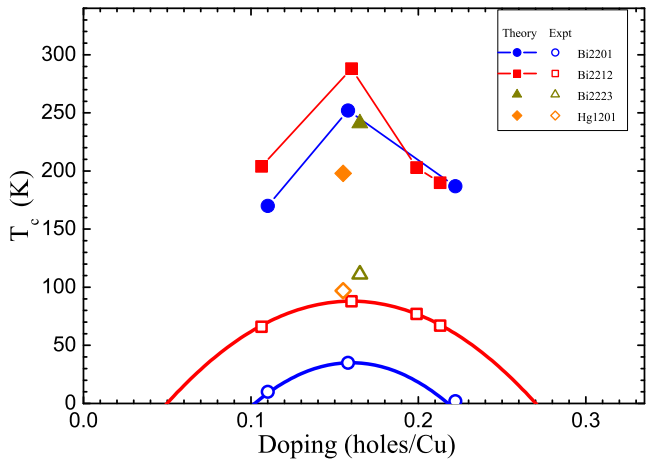}
\caption{Experimental critical temperature (open symbols) and $T_c$'s
calculated in the Eliashberg formalism (closed symbols) using the experimentally
measured \glue of Fig. \ref{fig_glue} at 290 Kelvin as input
parameters. }
\label{fig_tc}
\end{figure}
We take this analysis a step further by calculating $T_c$ from
the glue spectra below 100 meV ($\tilde{\Pi}_{pk}$ ) and above
100 meV ($\tilde{\Pi}_{cnt}$). The resulting coupling constants
and $T_c$'s are indicated in Table \ref{table1}. On the
underdoped side $\tilde{\Pi}_{cnt}$ vanishes, and $T_c$ is
given only by the coupling to the intense 50-60 meV peak in
$\tilde{\Pi}(\omega)$, but in this doping range we have to be
careful with the interpretation of our results. As mentioned in
the introduction our \glue spectra represent \textit{effective}
coupling functions, which may contain effects arising from
features not captured by the strong coupling equations
\ref{sigma} and \ref{Kubo}. If, for example, a pseudogap opens
in the electronic spectrum this will affect the shape of
$\tilde{\Pi}(\omega)$. These effects likely play a role for the
underdoped samples, but are not expected to affect much the
room temperature values, indicated in Table \ref{table1} and
Fig. \ref{fig_tc}. On the contrary, the larger temperature
dependence seen for underdoped samples in Fig. \ref{fig_glue}
may well be a result of the opening of a pseudogap. For the
overdoped samples the $T_c$'s calculated from
$\tilde{\Pi}_{pk}$ are {\em smaller} than the experimental
values. For example, for Bi2212 with the highest doping
$\tilde{\Pi}_{pk}$  gives only $T_c < 20$K, whereas
$\tilde{\Pi}_{cnt}$ gives 160 K, implying that the
glue-function above 100 meV is of crucial importance for the
pairing-mechanism. Since only electronic modes can have such
high energies, an important contribution to the high $T_c$
mechanism comes apparently from coupling to electronic degrees
of freedom, {\em i.e.}
spin\cite{scalapino-PRB-1986,millis-PRB-1990,maier-PRL-2008,norman-PRB-2006}
or orbital current fluctuations\cite{varma-PRL-1989}.

\section{Summary}\label{conclusion}
In summary, the \glue spectrum obtained from the optical
spectra of 10 different compounds using a strong coupling
analysis, is observed to consist of two features: (i) a robust
peak in the range of 50 to 60 meV and (ii) a doping dependent
continuum extending to 0.3 eV for the samples with the highest
$T_c$. We perform an important test of the internal consistency
of the strong coupling formalism by showing that the
temperature dependence of the optical spectra is determined by
Fermi and Bose factors in the strong coupling expressions. The
remaining temperature dependence of \glue can therefore be
taken to indicate that part of the \glue spectrum is electronic
in origin. We observe an intriguing correlation between the
doping trend of the experimental glue spectra and the critical
temperature. Finally we obtain an upper limit to the
contribution of electron-phonon coupling to the pairing of the
overdoped samples, which is too small to account for the
observed critical temperature.

\section{Acknowledgments}
We gratefully acknowledge C.M. Varma, D.J. Scalapino, J.
Zaanen, A.V. Chubukov, C. Berthod, J.C. Davis, and A. Millis for
stimulating discussions. This work is supported by the
Swiss National Science Foundation through Grant No.
200020-113293 and the National Center of Competence in
Research (NCCR) "Materials with Novel Electronic
Properties—MaNEP".

\section{Appendix}
\begin{figure}[h]
\centering
\includegraphics[width=8.5 cm]{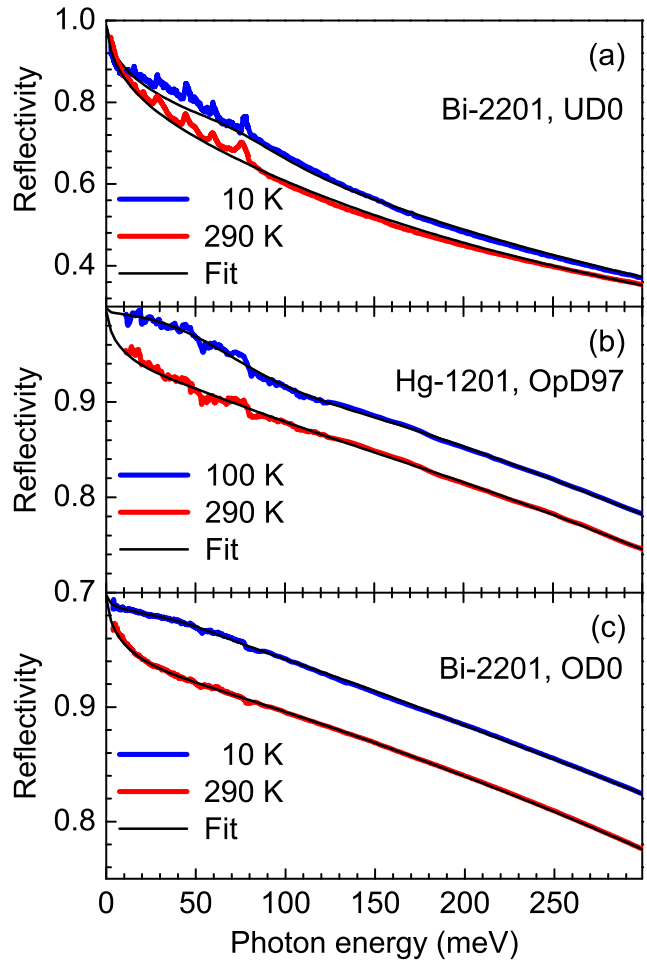}
\caption{Experimental reflectivity (red and blue lines) and fit curves (black lines) for selected samples and temperatures.
(a): underdoped non-superconducting Bi-2201 (UD0). (b): optimally doped Hg-1201 with $T_{c}\approx$ 97 K (OpD97) [\onlinecite{heumen-PRB-2007}]
(c): overdoped non-superconducting Bi-2201 (OD0). Weak sharp peaks, particularly visible for the strongly underdoped sample in panel (a) are due to transverse optical phonons, which we do not intend to fit.}
\label{fig_fitquality}
\end{figure}
The inversion of Eq's 1 and 2 allows to extract \glue from
experimental data of the optical conductivity, or related
optical spectra. The accuracy of the resulting \glue spectrum
is in practice limited by the convolution with thermal factors
expressed by Eq's 1 and 2 \cite{dordevic-PRB-2005}. Microscopic
models giving roughly the same \glue spectra, which differ
however in the details of the frequency dependence of this
quantity, may therefor provide fits to the directly measured
optical quantities, such as infrared reflectance spectra, which
at first glance look satisfactory, but the remaining
discrepancies with the experimental spectra may nevertheless be
of significant importance for the proper understanding of the
optical data. It is therefore of crucial importance to test the
'robustness' of each fit with regard to the spectral shape of
the \glue function imposed by such models. This robustness can
be tested by including in the fit-routine one or several
'oscillators' superimposed on the model function. When the
model glue function provides a complete description of the
electronic structure, adding extra oscillators will not result
in an improvement of the quality of the fit. We have used this
approach to test functional forms commonly used in the
literature, in particular the marginal Fermi liquid (MFL) model
\cite{varma-PRL-1989} and the Millis-Monien-Pines (MMP)
representation of the spin fluctuation
spectrum\cite{millis-PRB-1990}.

\begin{figure*}[!t] \centering
\includegraphics[width=17 cm]{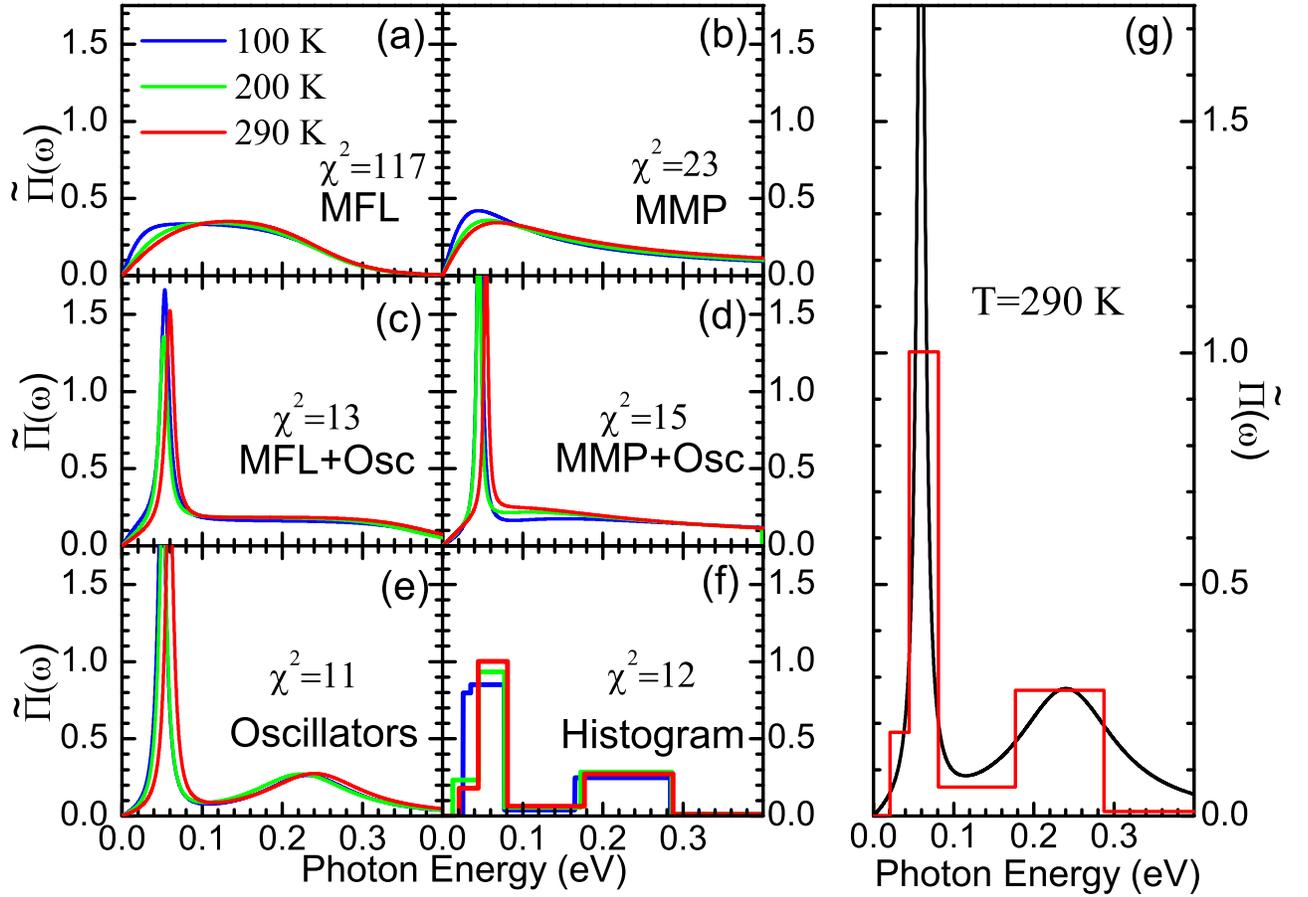}
\caption{Comparison of several models. The number of data
points over which $\chi_j^2$ is summed (see Eq. 1) is $N=400$.
The quoted values of $\chi^{2}$ are those for the room
temperature spectra. At 100 K these values increase by a factor
1.5 (MFL) and 3.5 (MMP). In contrast, in the oscillator and the
histogram model the $\chi^{2}$ was found to be independent of
temperature.} \label{fig_models}
\end{figure*}
We found that neither of these functional forms describe
completely the experimental data. In search of a more flexible
form of \glue we used a superposition of lorentzian oscillators
and found that it could be used to describe all available
experimental data in a consistent manner. The resulting \glue
functions and trends are equivalent to those in Fig.
\ref{fig_glue}. From these initial tests we concluded that due
to the thermal smearing expressed by Eq's 1 and 2 our \glue
spectra can only be determined with limited resolution. This
lead us to the use of a histogram representation, where each
block in the histogram represents a likelihood to find coupling
to a mode with a well determined coupling strength. For
the lowest frequency interval ($0 < \omega < \omega_1$) a
triangular shape was used instead of a block, which is
necessary to avoid problems with the convergence of the
integral $\lambda=2\int_0^{\infty} \omega^{-1}\tilde\Pi(\omega)
d\omega$. In practice the output generated by the fitting
routine has low intensity in this first interval, and the
triangles are therefore difficult to distinguish in Fig.
\ref{fig_glue}.

To give an example: The block centered at 55 meV seen in the
Hg-1201 sample in Fig. \ref{fig_glue} has $\lambda \sim 1$ and
a width of about 30 meV. Our histogram representation implies
the presence of a coupling to one or several modes between 45
meV and 75 meV with an integrated coupling strength of 1. The
histograms thus constitute the most detailed representation of
\glue given the precision of our experimental reflectivity and
ellipsometry spectra.

Examples of experimental reflectivity data together with the
fits are shown in Fig. \ref{fig_fitquality} for a selection of
representative data sets spanning the entire doping and
temperature range. As the fitted curves are within the limits
of the experimental noise, further reduction of $\chi^2$, while
in principle possible by fitting the statistical noise of the
data, can not improve the accuracy of the \glue functions.

Starting from a \glue  function we can calculate the optical
conductivity, which in turn is fed into standard Fresnel
expressions to calculate the experimentally measured
quantities, {\em i.e.} reflectivity and ellipsometric
parameters. The fitting routine is based on the
Levenberg-Marquardt algorithm and uses analytical expressions
for the partial derivatives of the reflectivity coefficient
$R$, and the ellipsometric parameters $\psi$ and $\Delta$
relative to the parameters describing the \glue function. The
algorithm is based on minimizing a functional $\chi^{2}$
which is given by,
\begin{equation}
\chi^{2}=\sum_{i=1}^{N}\left(\frac{R(\omega_{i})-f(\omega_{i},p_{1},...,p_{n})}{\sigma_{i}}\right)^{2}
\end{equation}
where $R(\omega_{i})$ is an experimentally measured datapoint,
$f(\omega_{i},p_{1},...,p_{n})$ is the calculated value in this
point based on parameters $p_{1},...,p_{n}$ and the difference
between these two is weighed by the errorbar $\sigma_{i}$
determined for $R(\omega_{i})$. For a given set of reflectivity
and ellipsometry data at one particular temperature, using a
standard PC, the iteration takes about 3 hours until
convergence is reached. The Levenberg-Marquardt least squares
method is an extremely powerful method to find the minimum of
$\chi^{2}$ in a multidimensional parameter space. To ensure
that $\chi^{2}$ has converged to the global minimum in
parameter space several tests have been performed, for each
individual sample and temperature displayed in Fig.
\ref{fig_models}, where in each test the optimization process
was started from a different set of starting parameters. To
give some idea of the robustness of our method we will here
discuss one representative example: optimally doped Hg-1201.

The models are evaluated based on the minimum found for
$\chi^{2}$. A comparison of Fig. \ref{fig_models} (a-d) shows
that the MMP model describes better the optical data then the
MFL model but that they give similar results if we add an extra
oscillator to these models. Panels \ref{fig_models} (e-f) show
the model independent results mentioned above and are very
similar to the modified MMP and MFL model. The models in these
last two panels have the same $\chi^{2}$ and the comparison in
Fig. \ref{fig_models}g shows that the histogram representation
realistically expresses the uncertainty in the position of the
low energy peak, while the correspondence between the features
in both models remains excellent. It is interesting that the
model with two oscillators is described by 6 parameters, while
the histogram representation uses 12 parameters. The fact that
the fit-routine adjusts the latter 12 parameters in such a
manner as to reproduce the two oscillators, proves that the
features represented in the righthand panel of Fig.
\ref{fig_models} are realistic.

\begin{figure}[!t]
\centering
\includegraphics[width=8.5 cm]{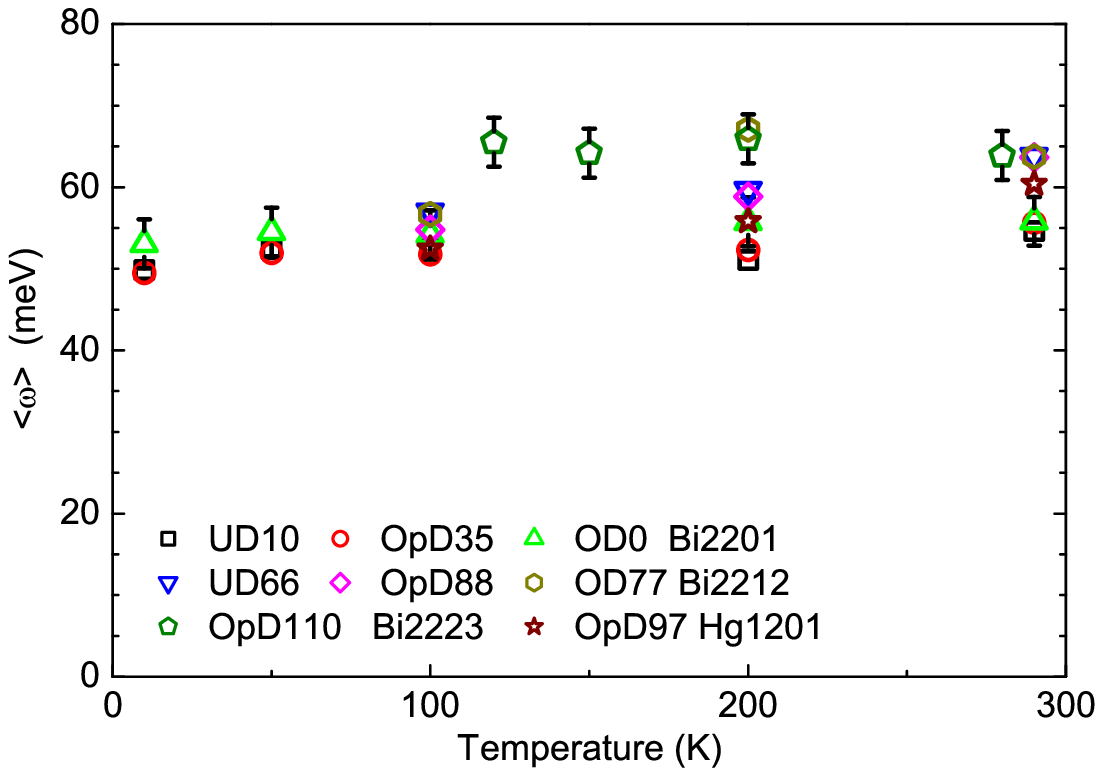}
\caption{Temperature dependence of $\left<\omega\right>$ for the \glue spectra presented in figure \ref{fig_glue}.
The error bars are defined in Eq. \ref{variance}.}
\label{fig_moment}
\end{figure}
The models presented in figure \ref{fig_models} allow us to
make an estimate of the uncertainty in the determination of the
frequency of the low energy peak. We define the first moment of
this peak as,
\begin{equation}
\left<\omega\right>=\int_0^{100 meV}\omega\tilde{\Pi}(\omega)d\omega \left/ \int_0^{100 meV}\tilde{\Pi}(\omega)d\omega. \right.
\end{equation}
The variance of $\left<\omega\right>$ is defined as,
\begin{equation}\label{variance}
\sigma^{2}=\frac{1}{N}\sum_{i=1}^{N}\left(\left<\omega_{i}\right>-\bar{\left<\omega\right>} \right)^{2}
\end{equation}
with $\bar{\left<\omega\right>}$ the mean of the moments of the
spectra presented in figure \ref{fig_models} and $i=1...N$ runs
over the number of spectra used ($N=6$ for each temperature).
For the Hg1201 room temperature spectra presented in figure
\ref{fig_models}a-f we find $\left<\omega\right>\approx$ 60 meV
and $\sigma\approx$ 3 meV. This value is approximately the same
for all samples. In figure \ref{fig_moment} we present the
temperature dependence of the first moment of the glue
functions presented in figure \ref{fig_glue}.


\begin{thebibliography}{99}
\bibitem{timusk-footnote}
    An alternative frequently used notation for this memory function\cite{timusk4} is $-2\Sigma^{op}(\omega)\equiv\hat{M}(\omega)$.
\end{thebibliography}
\end{document}